\documentclass[aps,prd,preprint,showpacs,nofootinbib,showkeys,superscriptaddress,floatfix]{revtex4-1}

\pdfoutput=1  
\usepackage{soul}
\usepackage{amsmath}
\usepackage{amssymb}
\usepackage{multirow}
\usepackage[american]{babel}
\usepackage{booktabs}
\usepackage{dcolumn}  
\usepackage[T1]{fontenc}  
\usepackage{graphicx}  
\usepackage[]{hyperref}  
\usepackage[utf8]{inputenc}  
\usepackage{url}
\usepackage[dvipsnames]{xcolor}
\usepackage{float}
\usepackage{subfigure}
\DeclareUnicodeCharacter{2212}{-}

\usepackage{tikz}

\usetikzlibrary{arrows,shapes,positioning,shadows,trees}

\tikzset{
	basic/.style  = {draw, text width=2cm, drop shadow, font=\sffamily, rectangle},
	root/.style   = {basic, rounded corners=2pt, thin, align=center,
		fill=green!30},
	level 2/.style = {basic, rounded corners=6pt, thin,align=center, fill=green!60,
		text width=8em},
	level 3/.style = {basic, thin, align=left, fill=pink!60, text width=6.5em}
}

\usetikzlibrary{arrows,decorations.pathmorphing,backgrounds,fit,positioning,shapes.symbols,chains}

\newcolumntype{d}[1]{D{.}{.}{#1}}
\newcolumntype{v}[1]{D{,}{,\ }{#1}}

\makeatletter

\newcommand{\Rmnum}[1]{\expandafter\@slowromancap\romannumeral #1@}

\makeatother

\usepackage{hyperref}
\hypersetup{
	colorlinks=true, 
	linkcolor=blue, 
	citecolor=magenta}

\begin{document}

	\title{Analysis of the action-based constrained adiabatic particle production model}
	
	\author{Souvik Pramanik}
	\email{souvick.in@gmail.com}
	\affiliation{Department of Physics, Ariel University, Israel}
	
	
	
	\begin{abstract}
		The theory of gravitationally induced adiabatic particle production offers an alternative approach to exploring the accelerating expansion of the universe. However, existing models generally lack a well-defined field-action formulation. This article aims to establish a suitable Lagrangian formulation consistent with the framework of particle creation. A general form of action is considered for this purpose. The corresponding field equations in comparison with those commonly used in the particle creation formalism leads to some specific form of the Lagrangian, which further constrains the form of the particle creation models. We perform the phase space stability analysis of the constrained model and compared with observational data. Our result shows that the constrained model is capable of explaining the cosmic evolution from the radiation  to the late-time cosmic acceleration if the model parameter lies within the range $1/3 < \beta < 2/3$.
	\end{abstract}
	\maketitle
	
	
	\section{Introduction}
	
	Observations from multiple independent astronomical missions \cite{snia1,snia2,Hinshaw:2012aka,Ade:2015xua} consistently confirm the accelerated expansion of the universe. Although understanding of this late time cosmic acceleration remains incomplete, two primary approaches are typically considered. The first one involves the introduction of a hypothetical dark energy component \cite{Copeland:2006wr} within the framework of Einstein's general relativity. On the other hand, modified theories of gravity is proposed \cite{Nojiri:2010wj,Capozziello:2011et,Nojiri:2017ncd, Cai:2015emx}, those include modifications of Einstein's theory.
	
	In addition to the above standard approaches, an alternative framework originated from the gravitationally induced adiabatic matter creation, a non-equilibrium thermodynamical process, has created a massive attention to the scientific community. The beginning of this theory was initiated by the seminal work of Schr\"{o}dinger's ideas \cite{Sch1939} and later was developed by several scientists including Parker and his collaborators~
	~\cite{Parker1968, Parker1969, Parker1970, Parker1977, Birrell1980, Birrell1982} and Zeldovich and others in Russia \cite{ZS1972, ZS1977, Grib1974, Grib1976, Grib1994}. However,  Prigogine et al. \cite{Prigogine1989} made a heuristic contribution by inserting the particle creation into the Einstein's field equations through the modification of the conservation equation as 
	\begin{equation}
		N_{;\mu}^{\mu}\equiv\mathit{n}_{,\mu}u^{\mu}+\Theta\mathit{n}=\mathit{n}%
		\Gamma~\Longleftrightarrow~N_{,\mu}u^{\mu}=\Gamma N,\label{balance-eqn}%
	\end{equation}
	where $\Gamma$ stands for the rate of change of the particle number in a  physical volume $V$ containing $N$ number of particles, $N^{\mu}%
	=\mathit{n}u^{\mu}$ represents particle flow vector; $u^{\mu}$ is the usual particle four velocity; $\mathit{n}=N/V$ is the particle number density, $\Theta=u_{;\mu}^{\mu}$, denotes the fluid expansion. The newly introduced quantity $\Gamma$ is of special significance: it represents the rate of particle production. Though, its precise form is still unclear, a constraint on $\Gamma$ arises from the validity of the generalized second law of thermodynamics which requires $\Gamma\geq0$. 
	But, the nature of the particles created by the gravitational field is completely unknown raising the question about the characteristic and physical properties of such particles.
	Although the types of particles that can be produced have been shown to be significantly constrained by local gravity tests \cite{Ellis1989, Hagiwara2002, PR2003}.
	Practically, radiation has no significant effect on the late-time accelerated expansion of the universe. In contrast, dark matter is one of the dominant sources after the unknown \textquotedblleft dark energy\textquotedblright\ component. What follows is that, we may assume that the particles produced by this gravitational field are simply the cold dark-matter particles. Based on this assumption, it has been argued that cosmological models incorporating different particle creation rates can successfully mimic the $\Lambda$CDM cosmology \cite{SSL2009, LJO2010, LGPB2014, FPP2014, CPS2015}. In particular, a constant rate of particle creation has been shown to account for the Big Bang singularity, intermediate cosmological phases, and the eventual approach to a de-Sitter regime \cite{HP2015}. Furthermore, Nunes and Pav\'{o}n \cite{NP2015} demonstrated that matter creation models can explain the phantom-like behavior of the Universe --evident in observations \cite{Planck2014, Rest2014, XLZ2013, CH2014, SH2014} without the need of any phantom fields \cite{Caldwell2002}. The broader cosmological implications of matter creation models that exhibit phantom behavior have also been explored in subsequent work \cite{NP2016}. Furthermore, particle production mechanisms have attracted growing attention in the context of modified gravity theories \cite{Harko2015,CLP2016}, and subsequent works \cite{GS1979, Varun, Desai2016,Nunes2016}.
	
		Although, there are a few $f(R)$ gravity models which can study particle production scenarios in a well defined way \cite{Pinto2022, Calvao1992, Arbuzova2013, Kolb2024, SotiriouFaraoni2010}, but in many cases, the particle production rate $\Gamma$ has been chosen hypothetically, and inserted into the acceleration and continuity equations \cite{CPS2015, CardenasCruz2024, Lima1996, Komatsu2020,  Abramo1996, ZimdahlPavon1993}. 
	In any gravitational framework, whether Einstein gravity or a modified gravity theory, the field equations are typically obtained from the variation of an action with respect to the metric and matter fields. Motivated by this foundational thought, in the present work we aim to formulate a Lagrangian description for particle creation models. Indeed, there is a classical Lagrangian (\ref{lag-1}) which reproduces the standard Friedmann and continuity equations. To incorporate the particle creation mechanism, we will consider a generalization of that which includes arbitrary higher-order derivatives of the scale factor and scalar field (see Section-\ref{III}). We will derive the corresponding field equations and compare them with those commonly used in the particle creation formalism. We will show that this comparison leads to formulate the modified Lagrangian consistent with particle creation dynamics. Additionally, consistency of the field equations will impose constraint on the allowed forms of particle creation models that we believe have not been taken care of previously. Thereafter, we will perform a phase-space stability analysis for the model restricted by our Lagrangian framework in Section-\ref{IV} and explain the cosmological eras with some specific bounds of the model parameter. Finally, we will perform numerical simulations for the constrained model in Section-\ref{V} to show compatibility with observed data.   
	
	\section{Basic formalism}
	\label{sec-basics}
	
	According to observational evidence, the geometrical configuration of our universe is well described by the spatially flat Friedmann-Lema\^{i}tre-Robertson-Walker (FLRW) line element
	\begin{equation}
		ds^2= -dt^2+a^2(t) (dx^2+ dy^2+dz^2), \label{flrw}
	\end{equation}
	where, $a(t)$ is the scale factor of the expanding universe.  In this background, Einstein's field equations are written as (considering $8 \pi G = 1$)
	\begin{eqnarray}
		3 H^2 &=&  \rho,\label{EF1}\\
		2\dot{H}+ 3 H^2 &=& -  p,\label{EF2}\\
		\dot{\rho} + 3 H (\rho+ p) &=& 0, \label{con}
	\end{eqnarray}
	where, the overhead dots represent differentiation with respect to the cosmic time $t$, $\rho$ and $p$ are the total energy density and pressure of matter, respectively. $H= \dot{a}/a$ is the FLRW Hubble parameter. Any two of the above equations (\ref{EF1}, \ref{EF2}, \ref{con}) are mutually independent, means the unchosen one can be derived using the others. The equation (\ref{con}) is known as the conservation equation of the matter, can also be derived using Bianchi's identity. If the background fluid is dominated by a scalar field $\phi(t)$, the energy density $\rho$ and pressure $p$ are calculated from the temporal and special components of the energy-momentum tensor $T^{\mu\nu}$ as
	\begin{equation}
		\rho = \frac{1}{2} \dot{\phi}^2 + V(\phi), \ \ p = \frac{1}{2} \dot{\phi}^2 - V(\phi), \label{rho-p}
	\end{equation}
	respectively. The above field equations (\ref{EF1},\ref{EF2},\ref{con}) with the above density and pressure terms (\ref{rho-p}) can be achieved from a classical action of the form 
	\begin{eqnarray}
		S = \int \mathcal{L}_0  (a, \dot a, \phi, \dot \phi) \ dt = \int \left[ 3 a \dot{a}^2 - \frac{1}{2} a^3 \dot{\phi}^2 + a^3 V(\phi) \right] \ dt.  \label{lag-1}
	\end{eqnarray}
		If we consider a quasi-spherical polar coordinate with zero curvature $k=0$, the geometry is described by the metric
		$$ds^2 = - N^2(t) dt^2 + a^2(t) [dr^2+ r^2 d\Omega^2],$$
		where $N(t)$  is the lapse function \cite{Vakili2014, Vakili2011, Halliwell1990}. Substituting the above metric into the Einstein- Hilbert action $\int d^4x \sqrt{-g} (R/2 + L_m)$ and integrating out the special volume integral, we obtain the standard point-like mini-superspace action of the  form \footnote{ One can evaluate $\sqrt{-g} = N a^3, g^{00} = 1/ N^2$}
		\begin{equation}
			S = \int dt\,N\!\left[\frac{1}{2N^2}G_{AB}(q)\dot q^A\dot q^B - U(q)\right], \label{L0}
		\end{equation}
		where, $q^A =(a,\phi)$ and $G_{AB} = diag(6a,-a^3)$. Expanding the terms of (\ref{L0}) for $N=1$ provides the Lagrangian (\ref{lag-1}). The evolution and continuity equations (\ref{EF2}, \ref{con}) and the Friedmann equation (\ref{EF1}) (letting the Hamiltonian constraint to zero for energy conservation) can also be obtained by varying the action (\ref{L0}) with respect to the variables $q^A$ and $N$, respectively. The action (\ref{L0}) is in a mini-superspace of $4D$ and so the concept of general covariance does not applicable in this case \cite{Vakili2014, Vakili2011, Halliwell1990}. However, this action is invariant under time reparametrization given by $t\mapsto \bar t (t)$, or in other words invariant under a general coordinate transformation in $t$. In addition, for (\ref{L0}) (or equivalently (\ref{lag-1})) one can define a vector field $\hat X$ by
		\begin{equation}
			\hat X = \alpha^A({\bf q}) \frac{\partial}{\partial q^A} + \frac{d\alpha^A(\bf q)}{dt} \frac{\partial}{\partial \dot q^A} \label{vec-x}
		\end{equation}
		on the tangent space $TQ=(\bf q, \bf \dot q)$, where $\alpha^A(\bf q)$s are unknown functions on configuration space, so that, $\hat X$ generates a Noether symmetry if $\hat X \mathcal{L}_0 =0$. If one uses the Euler-Lagrange equations $\frac{dP_A}{dt}= \frac{\partial \mathcal{L}_0}{\partial q^A}$, where $P_A= \frac{\partial \mathcal{L}_0}{\partial \dot q^A}$ then conserved quantity can be found to be $Q = \alpha^A({\bf q}) P_A$ \cite{Vakili2014, Vakili2011, Halliwell1990}.

	\section{Field action for particle creation models} 
	\label{III}
	
	In the presence of continuous particle creation driven by the time varying gravitational field, the field equations modify to
	\begin{eqnarray}
		3 H^2&=& \rho,\label{EF- p -1}\\
		2\dot{H}+ 3 H^2 &=& - (p + p_c),\label{EF - p - 2}\\
		\dot{\rho} + 3 H (\rho+ p + p_c) &=& 0. \label{con-eqn}
	\end{eqnarray}
	In (\ref{EF - p - 2}, \ref{con-eqn}), the new quantity $p_c$ represents the additional pressure term responsible for the continuous creation of matter. If one focused on the `adiabatic condition' that the entropy per particle is constant, the extra pressure term $p_c$ is assumed to be a very simple form
	\begin{eqnarray}
		p_c&=& -\frac{\Gamma}{3H} (p+\rho), \label{pc}
	\end{eqnarray}
	where $\Gamma$ is the rate of particle creation due to the time-varying gravitational field generally considered as a function of time, the scale factor and its derivatives $ \Gamma \equiv \Gamma(t, a(t), \dot a(t),\ddot a(t),...)$. In principle, $\Gamma> 0$ represents matter creation, in contrary, $\Gamma< 0$ means matter annihilation and $\Gamma= 0$ is the case of no particle production which reproduces the field equations (\ref{EF1},\ref{EF2},\ref{con}). If the particle are produced due to interaction between the scalar field and metric sector, then using (\ref{rho-p}) we have
	\begin{equation}
		p_c = -\frac{\Gamma}{3 H} \dot \phi^2 = -\frac{ \dot \phi^2}{3 H} \times \Gamma( t, a, a^{(1)},a^{(2)},...,a^{(n)}), \label{p-c}
	\end{equation} 
	where, ${(i)}$ is the $i^{\rm th}$ derivative with respect to the cosmic time $t$. Substituting (\ref{p-c}) into the above Friedmann equations (\ref{EF- p -1}, \ref{EF - p - 2},\ref{con-eqn}), we have 
	\begin{eqnarray}
		&& 3 H^2 = \left( \frac{1}{2} \dot{\phi}^2 + V(\phi) \right),   
		\label{EF-p-1-1}\\
		&& 2\dot{H}+ 3 H^2 = -  \left( \frac{1}{2} \dot{\phi}^2 - V(\phi) - \frac{\dot \phi^2}{3H} \ \Gamma(t, a, a^{(1)}, ..., a^{(n)}) \right),\label{EF-p-2-2}\\
		&& \ddot{\phi} + V_\phi + 3 H \dot \phi - \dot \phi \times \Gamma(t, a, a^{(1)}, ..., a^{(n)}) ) = 0. \label{con-eqn-1}
	\end{eqnarray}
	
		The above system (\ref{EF-p-1-1} - \ref{con-eqn-1}) has been widely studied for a few decades \cite{CardenasCruz2024, Lima1996, Komatsu2020,  Abramo1996, ZimdahlPavon1993}. However, there does not exist a proper field action which can achieve it. Therefore, for the completeness of the above system, we argue here that it should has to be yield at least from the generalization of the point-like mini-superspace Lagrangian of the form
		\begin{eqnarray}
			S &=& \int \mathcal{L}(a, a^{(1)},a^{(2)},...,a^{(n)}, \phi, \phi^{(1)},\phi^{(2)},...,\phi^{(n)}) dt, \nonumber\\
			\text{with} \ \mathcal{L} &=& \mathcal{L}_0 \times g(a, a^{(1)},a^{(2)},...,a^{(n)}, \phi, \phi^{(1)},\phi^{(2)},...,\phi^{(n)}),
			\label{act-22}
		\end{eqnarray}   
		for some suitable choice $g$ of background fields, where $\mathcal{L}_0$ is given by (\ref{lag-1}). The above form of Lagrangian is the most general choice \footnote{without taking care of the symmetry like time reparametrization because of first at least find a suitable Lagrangian.}, because of $g$ can be contributed as an additive or multiplicative term to the usual Lagrangian. Additionally, as $g$ is a function of both the fields $a$, $\phi$ and their time derivatives, we can think of our choice a most general one and compatible with any arbitrary particle creation model. 
	
	For the continuity equation, the general form of the Euler-Lagrange equation with respect to the variable $\phi$ provides
	\begin{eqnarray}
		&& \ddot \phi + V_\phi +\frac{3 \dot \phi \dot a}{a} + \frac{\dot g \dot \phi}{g}+ \frac{1}{g}\frac{\partial g}{\partial \phi} \left( \frac{3 \dot a^2}{a^2} -\frac{\dot \phi^2}{2} + V\right) \nonumber \\ 
		&& - \frac{1}{g} \frac{\partial g}{\partial \dot \phi}\left(\frac{6 \dot a \ddot a}{a^2} +\frac{3\dot a^3}{a^3} - \frac{3\dot a \dot \phi^2}{2 a} - \dot \phi \ddot \phi +\frac{3\dot a V}{a} + \dot \phi V_\phi \right)  \nonumber\\
		&& - \frac{1}{g} \frac{d}{dt} \left(\frac{\partial g}{\partial \dot \phi} \right)\left( \frac{3\dot a^2}{a^2} - \frac{\dot \phi^2}{2 } + V \right) + \frac{1}{g a^3}\sum_{i=2}^{n} (-1)^i \frac{d^i}{dt^i}\left( \frac{\partial L}{\partial \phi^{(i)}} \right) =0 . \label{phi-eq-1}
	\end{eqnarray} 
	Comparing (\ref{phi-eq-1}) and (\ref{con-eqn-1}), we can conclude that $g$ cannot be a function of $\phi$ and it's time derivatives \footnote{because of the presence of potential $V$ in all the extra terms in (\ref{phi-eq-1}) incompatible with $\Gamma(t, a, a^{(1)}, ..., a^{(n)})$.}, i.e 
	\begin{equation}
		\frac{\partial g}{\partial \phi} =0, \ \frac{\partial g}{\partial \phi^{(i)}} =0.
	\end{equation}
	Therefore $g\equiv g(t, a, a^{(1)},...a^{(n)})$, and a further comparison of the fourth term of (\ref{phi-eq-1}) to the particle creation term in (\ref{con-eqn-1}) provides
	\begin{equation} 
		\frac{\dot g}{g} = - \Gamma(t, a, a^{(1)}, ..., a^{(n)}) \label{first-con}
	\end{equation}
	Similarly, the generalized Euler-Lagrange equation for the variable $a$, provides
	\begin{eqnarray}
		\frac{2 \ddot a}{a} + \frac{\dot a^2}{a^2} &=& - \left(\frac{\dot \phi^2}{2} -V(\phi)\right)  - \frac{2 \dot a \dot g}{a g} + \frac{a}{ g}\frac{\partial g}{\partial a} \left(\frac{\dot a^2}{a^2} - \frac{\dot \phi^2}{6} +\frac{V}{3} \right)  \nonumber\\
		&& - \frac{a}{3 g}\frac{d}{dt}\left(\frac{\partial g}{\partial \dot a}\right) \left(\frac{3 \dot a^2}{a^2} - \frac{\dot \phi^2}{2} + V \right) \nonumber\\
		&& + \frac{1}{3 g a^2} \frac{\partial g}{\partial \dot a} \frac{d}{dt} \left( 3 a \dot a^2 - 1/2 a^3 \dot \phi^2 + a^3 V \right) \nonumber\\
		&& - \frac{1 }{3 a^2 g} \sum_{i=2}^{n} (-1)^i \frac{d^i}{dt^i}\left(\left[3 a \dot{a}^2 - \frac{1}{2} a^3 \dot{\phi}^2 + a^3 V(\phi)\right]\frac{\partial g}{ \partial a^{(i)}}\right)  . \label{a-eq}
	\end{eqnarray} 
	Comparing (\ref{a-eq}) and the second Friedmann equation (\ref{EF-p-2-2}) we have \footnote{for the same resign as above}
	\begin{equation}
		\frac{\partial g}{\partial a^{(i)}} =0 \ for\ i=1,2,...,n.
	\end{equation}
	This further restricts the function $g$ to $g\equiv g(t,a)$ only, and we are left with
	\begin{equation}
		\frac{2 \ddot a}{a} + \frac{\dot a^2}{a^2} = - \left(\frac{\dot \phi^2}{2} -V(\phi)\right)  - \frac{2 \dot a \dot g}{a g} + \frac{a}{ g}\frac{\partial g}{\partial a} \left(\frac{\dot a^2}{a^2} - \frac{\dot \phi^2}{6} +\frac{V}{3} \right). \label{2-friedmann}
	\end{equation}
	At this point, we would like to emphasize that the first Friedmann equation (\ref{EF-p-1-1}) is consistent with the above restriction of $g$ if $g = g(a)$ only. The Hamiltonian constraint with the above form of $g$ is calculated as
	\begin{eqnarray}
		\mathcal{H} &=& \dot a \frac{\partial L}{\partial \dot a} + \dot \phi \frac{\partial L}{\partial \dot \phi} - \mathcal{L} \nonumber\\
		&=& g \times \left( 3 a \dot a^2  - \frac{a^3}{2} \dot{\phi}^2 - a^3 V(\phi) \right).
	\end{eqnarray}
	Considering $ \mathcal{H}$ to be zero the first Friedmann equation (\ref{EF-p-1-1}) is reproduced. Substituting this equation and the above form of $g$ into (\ref{2-friedmann}), we have
	\begin{eqnarray}
		\frac{2 \ddot a}{a} + \frac{\dot a^2}{a^2} &=& - \left(\frac{\dot \phi^2}{2} -V \right)  - \frac{2 \dot a^2}{a g} \frac{\partial g}{\partial a} + \frac{a}{ 3 g}\frac{\partial g}{\partial a} \left(\frac{3 \dot a^2}{a^2} - \frac{\dot \phi^2}{2} + V  \right) \nonumber\\
		&=& - \left(\frac{\dot \phi^2}{2} -V \right)  -  \frac{a}{ 3 g}\frac{\partial g}{\partial a} \left(\frac{3 \dot a^2}{a^2} + \frac{\dot \phi^2}{2} - V  \right) \nonumber\\ 
		&=& - \left(\frac{\dot \phi^2}{2} -V \right)  - \frac{a}{ 3 g}\frac{\partial g}{\partial a} \dot \phi^2 \label{FR-3}
	\end{eqnarray}
	Again, comparing (\ref{FR-3}) with above acceleration equation (\ref{EF-p-2-2}), we finally have 
	\begin{equation}
		\frac{a}{g}\frac{\partial g}{\partial a} = - \frac{\Gamma}{H} \implies \frac{\dot a}{g}\frac{\partial g}{\partial a} = - \Gamma
		\label{seco-con}
	\end{equation}
	This relation is fairly consistent with the above condition (\ref{first-con}) for the restriction of $g\equiv g(a)$, only \footnote{else $\dot g$ in equation (\ref{first-con}) will produce a lot of high order derivative terms containing $a^{(i)}$ incompatible with left hand side of (\ref{seco-con})}. With the restriction on $g$, (\ref{seco-con}) provides the allowed form of particle creation model to
	\begin{equation}
		\Gamma = \Gamma (a,\dot a) = \dot a \times h(a). \label{model} 
	\end{equation}      
	where $h(a)$ is an arbitrary function of $a$. Plunging back (\ref{model}) into (\ref{seco-con}) the relevant term of the Lagrangian found to be
	\begin{equation}
		g(a) = e^{ - \int h(a) da}, \label{ga}
	\end{equation}
	so that, the Lagrangian becomes 
	\begin{equation}
		\mathcal{L} = \left( 3 a \dot a^2 - \frac{a^3}{2} \dot \phi^2 + a^3 V(\phi) \right) e^{ - \int h(a) da}. \label{final-L}
	\end{equation}
		This is the only form of Lagrangian compatible with the system (\ref{EF-p-1-1} - \ref{con-eqn-1}) to study particle creation model, but with the model restriction evaluated in (\ref{model}). In other words, a point-like mini-superspace Lagrangian cannot exists for particle creation models, other than (\ref{model}). Below, we will concentrate on some symmetries of the above Lagraigian (\ref{final-L}).
		
	
	
		With the final form of Lagrangian (\ref{final-L}), the action can be written in the lapse form as
		\begin{equation}
			S = \int dt\, N\left\{\frac{1}{2N^2}\big( 6a\,\dot a^2 - a^3\dot\phi^2\big) + a^3 V(\phi)\right\} e^{-\int h(a) da}.
		\end{equation}
		If we consider a general coordinate transformation in $t$ of the form $\bar t = f(t)$, then $d\bar t = df/dt \times dt = f_t dt$. This provides $\dot a = a' f_t, \dot \phi = \phi' f_t$ and $N \ dt = N \ d\bar t / f_t = \bar N d\bar t$, where we defined $a'=da/d\bar t, \phi'=d\phi /d\bar t, \bar N = N/f_t$. Under the above transformation the $4D$ metric remains form invariant: $ds^2 = - \bar N^2 d\bar t^2 + a^2(\bar t) [dr^2+ r^2 d\Omega^2]$. Now, for the mini-superspace action we have  
		\begin{eqnarray}
			S &=& \int \frac{d \bar t}{f_t} \bar N f_t \left\{\frac{1}{2 \bar N^2 f^2_t}\big( 6a\,a'^2 f^2_t - a^3 \phi'^2 f^2_t\big) + a^3 V(\phi)\right\} e^{-\int h(a) da} \nonumber \\
			&=& \int d \bar t \ \bar N \left\{\frac{1}{2 \bar N^2}\big( 6a\,a'^2 - a^3 \phi'^2 \big) + a^3 V(\phi)\right\} e^{-\int h(a) da}.
		\end{eqnarray}
		Clearly, the above action remains invariant since the extra quantity $g(a) = e^{-\int h(a) da}$ is restricted to a function of $a$ only, not its derivatives. This finding further strengthen the restriction to the form of $g$. We can also find the invariant quantity using the Noether symmetry stated above in Section-\ref{sec-basics}. Since, the momentum in this case is $P_A= \frac{\partial L}{\partial \dot q^A} = (6 a \dot a , - a^3 \dot \phi) e^{-\int h(a) da}$, the corresponding conserved quantity (Noether charge) can be evaluated as $Q= [6 a  \dot a \ \alpha(a,\phi)  - a^3 \dot \phi \ \beta(a,\phi) ] e^{-\int h(a) da}$, where the unknown functions $\alpha, \beta$ need to find from the Noether symmetry
		\[
		\hat X L=\alpha(a,\phi)\frac{\partial L}{\partial a}+\beta(a,\phi)\frac{\partial L}{\partial\phi}
		+\dot\alpha\frac{\partial L}{\partial\dot a}+\dot\beta\frac{\partial L}{\partial\dot\phi} =0,
		\]
		for different choices of $h(a)$. This study ensures that the final form of Lagrangian (\ref{final-L}) also produces the Noether symmetry.
	
	\section{Dynamical system and stability analysis}
	\label{IV}	
	
	With the allowed form of the particle creation model (\ref{model}), field equations are written as
	\begin{eqnarray}
		&& 3 H^2 = \left( \frac{1}{2} \dot{\phi}^2 + V(\phi) \right),   
		\label{EF-p-1-1-1}\\
		&& 2\dot{H}+ 3 H^2 = -  \frac{1}{2} \dot{\phi}^2 + V(\phi) + \frac{a\times h(a)}{3} \dot \phi^2,\label{EF-p-2-2-2}\\
		&& \ddot{\phi} + V_\phi + 3 H \dot \phi - \dot \phi \times \dot a h(a) = 0. \label{con-eqn-1-2}
	\end{eqnarray}	
	Before forming a autonomous system and analyzing it, we will introduce the contribution of  radiations and baryons into the above system. The total energy density and pressure of the system are then $\rho = \rho_\phi + \rho_r +\rho_b$ and $p = p_\phi + p_r + p_c$ (since $p_b =0$), respectively, whereas the contribution of total matter come from $\rho_\phi$ and $\rho_b$. Since we assume here that the particle creation pressure (\ref{p-c}) depends on $\rho_\phi, p_\phi$ and the metric $a$, only, consequently, the Lagrangian for radiation $\mathcal{L}_r$ and baryons $\mathcal{L}_b$ can be treated as additional quantities to the above Lagrangian (\ref{final-L}). Physically, this means that the process of particle creation is not affected by radiation and baryonic matter. In addition to this, we will consider the individual conservation law hold separately for each sectors, the radiation, baryons and scalar field. With all these assumptions in hand, the system to study is
	\begin{eqnarray}
		&& 3 H^2 =  \frac{1}{2} \dot{\phi}^2 + V(\phi) + \rho_r +\rho_b,   
		\label{d1}\\
		&& 2\dot{H}+ 3 H^2 = -  \frac{1}{2} \dot{\phi}^2 + V(\phi) + \frac{a\times h(a)}{3} \dot \phi^2 - p_r ,\label{d2}\\
		&& \ddot{\phi} + V_\phi + 3 H \dot \phi - \dot \phi \times \dot a h(a) = 0, \label{d3} \\
		&& \dot \rho_r + 4 H \rho_r =0, \label{d4} \\
		&& \dot \rho_b + 3 H \rho_b =0. \label{d5} 
	\end{eqnarray}	
	Beside the common dimensionless variables defined by 
	\begin{equation}
		x = \frac{\dot{\phi}}{\sqrt{6}H}, y = \frac{\sqrt{V(\phi)}}{\sqrt{3}H}, \Omega_r = \frac{\rho_r}{ 3 H^2},  \Omega_b = \frac{\rho_b}{ 3 H^2},
	\end{equation}
	here, additionally we will consider $z = a \times h(a)$ as a new variable responsible for particle creation process. With these definitions, the evolution equation (\ref{d1}), the effective equation of state $\omega_{eff}$ and the acceleration equation (\ref{d2}) turn out to be \begin{eqnarray}
		&& x^2 + y^2 + \Omega_r + \Omega_b = 1 \label{constrain} \\
		&& \omega_{eff} = \frac{p_\phi + p_r + p_c}{\rho_\phi + \rho_b + \rho_r} = x^2 - y^2 + \frac{\Omega_r}{3}  - \frac{2 z}{3} x^2, \label{omega-eff}\\
		&& \frac{\dot H}{H^2} = - \frac{3}{2} (1+ \omega_{eff}) \label{acc-eqn}.
	\end{eqnarray}
	Due to the Friedmann constraint (\ref{constrain}) we can drop out $\Omega_b$ and consider it as a parameter. The autonomous system to study can be formed as  
	\begin{eqnarray}
		x' &=& -3 x + z x + \frac{3x}{2} \left(1 + x^2 - y^2 + \frac{\Omega_r}{3} - \frac{2 z}{3}x^2\right) + \lambda \sqrt{\frac{3}{2}} y^2,  \label{x}\\
		y' &=& \frac{3}{2} y \left(1 + x^2 - y^2 + \frac{\Omega_r}{3} - \frac{2 z}{3}x^2\right) - \lambda \sqrt{\frac{3}{2}} x y , \label{y} \\
		\Omega_r' &=& \Omega_r \left( - 1 + 3 \left( x^2 - y^2 + \frac{\Omega_r}{3} - \frac{2z}{3} x^2 \right) \right) , \label{omega-r} \\
		z' &=& z + z_1 z^2 , \label{z} \\
		z_1' &=& z z_2 z_1^{3/2} - 2 z z_1^2 , \label{z-1} \\
		\lambda' &=& - \sqrt{6} (\Lambda - 1 ) \lambda^2 x ,\label{lambda}\\
		\Lambda' &=& \sqrt{6} x ( - \lambda \Lambda + \Lambda \Lambda_1 + 2 \lambda \Lambda^2) ,\label{Gamma}
	\end{eqnarray}
	where $`` \ ' \ " = d/dN$, with $N= ln(a)$. The other variables are $z_1 = h_a/h^2, z_2 = h_{aa}/ h_a^{3/2}, \lambda = - V_\phi / V , \Lambda = V V_{\phi\phi}/ V_\phi^2$, and $\Lambda_1 = V_{\phi\phi\phi} / V_{\phi\phi}$. Notice that, we include the dynamical equations for $z_1$ and $\Lambda$ in addition to the conventional equations. The reason is to investigate whether some suitable form of $h(a), V(\phi)$ can be found, such that the forms are compatible with their higher order evolution and can produce a critical point \footnote{one can further include dynamical equations for $z_2, \Lambda_1$, and so on, but here we see that those cannot produce critical points.}. 
	
	First, the equations (\ref{z},\ref{z-1}) are governed by variables $z, z_1,z_2$ only, so that critical points can be found independently. The critical point $z=0$ can be neglected as it does not show particle creation. The other conditions from (\ref{z},\ref{z-1}) are, respectively, 
	\begin{eqnarray}
		1+ z_1 z =0 , & \ \ \ \  & z_2 = 2 z_1^{1/2} \nonumber\\
		\implies  1+ a . \frac{h_a}{h} =0 , & \ \ \implies & \frac{h_{aa}}{h_a} = \frac{2 h_a}{h} \nonumber \\
		\implies  h(a) = \frac{c_1}{a} , &\ \ \implies & h(a) = - \frac{c_2}{a}
	\end{eqnarray}
	Clearly, the forms of $h(a)$ are consistent if $c_1 = - c_2$. So, for $h(a) = \frac{c_1}{a}$ the equations of $z', z_1'$ vanish without the critical point $z=0$. We are interested in this form. Because, no other form can consistently make the higher order evolutions vanishes while a critical point is calculate. The particle creation model is then $\Gamma = \dot  a \times c_1 / a \implies p_c = c_1/3 \ \dot \phi^2 = \beta \ \dot \phi^2$ (say), for which the additional variable becomes $z=3 \beta$. On the other hand, $x=0$ is a critical point of (\ref{lambda}, \ref{Gamma}). Otherwise, the right hand side vanishes if $[\lambda =0$ or $\Lambda =1]$ or $[\Lambda \Lambda_1 + 2 \lambda \Lambda^2 - \lambda \Lambda = 0]$ respectively, for the first and second equations. If $\lambda =0$ ($\implies V(\phi) = c_2$), we no longer have to consider (\ref{Gamma}). If $\Lambda=1$, the potential can be found as $V(\phi) = c_1 e^{- c_2 \phi} \implies \lambda = c_2$, which satisfies the condition obtained from the second equation. It should be noted that, other potentials cannot generate any critical point from (\ref{lambda}, \ref{Gamma}) other than $x=0$ \footnote{for instance, $V(\phi) = \phi^4$, the condition from second equation is satisfied but not the first one}, and so, in such cases, the critical points of the entire phase space will be calculated with $x=0$, only. With $z= 3 \beta$, the above autonomous system reduces to   
	\begin{eqnarray}
		x' &=& -3 x(1 - \beta) + \frac{3x}{2} \left(1 + x^2 - y^2 + \frac{\Omega_r}{3} - 2 \beta x^2\right) + \lambda \sqrt{\frac{3}{2}} y^2  \label{x-2}\\
		y' &=& \frac{3}{2} y \left(1 + x^2 - y^2 + \frac{\Omega_r}{3} - 2 \beta x^2\right) - \lambda  \sqrt{\frac{3}{2}} x y \label{y-2} \\
		\Omega_r' &=& \Omega_r \left( - 1 + 3 \left( x^2 - y^2 + \frac{\Omega_r}{3} - 2 \beta x^2 \right) \right). \label{omega-r-2} 
	\end{eqnarray}
	
	\subsection{Phase space analysis with $\lambda = constant \neq 0$: }
	
	The critical points are the solutions of \(x' = y' = \Omega_r' = 0\), which provides
	\begin{eqnarray}
		& -3 x(1 - \beta) + \frac{3x}{2} \left(1 + x^2 - y^2 + \frac{\Omega_r}{3} - 2 \beta x^2\right) + \lambda \sqrt{\frac{3}{2}} y^2 & = 0 \label{x-3} \\
		& \frac{3}{2} y \left(1 + x^2 - y^2 + \frac{\Omega_r}{3} - 2 \beta x^2\right) - \lambda  \sqrt{\frac{3}{2}} x y & = 0 \label{y-3} \\
		& \Omega_r \left( - 1 + 3 \left( x^2 - y^2 + \frac{\Omega_r}{3} - 2 \beta x^2 \right) \right) & = 0. \label{omega-3}
	\end{eqnarray}
	From both equations (\ref{y-3}, \ref{omega-3}), we have two possibilities:
	\begin{eqnarray}
		&& y = 0, \ or \ \frac{3}{2} \left(1 + x^2 - y^2 + \frac{\Omega_r}{3} - 2 \beta x^2\right) - \lambda  \sqrt{\frac{3}{2}} x   = 0 , \label{con-2} \\
		&& \Omega_r = 0, \ or \ -\frac{1}{3} + \left(x^2 - y^2 + \frac{\Omega_r}{3} - 2\beta x^2\right) = 0. \label{con-3}
	\end{eqnarray}
	\subsubsection{Case 1: \(y = 0\) and \(\Omega_r = 0\)}
	From the Friedmann constraint we have \( \Omega_b = 1 - x^2 \), whereas from (\ref{x-3}) we have
	\[ x \left( - 1 +  \beta + \frac{1}{2} + \frac{1}{2}x^2 (1 - 2 \beta )  \right) = 0 \]
	\[\implies x = 0 ,\ x = 1 , \ \beta = 1/2 .\]
	For \(x = 0\), the critical point is \(A = (0, 0, 0)\) with \(\Omega_b = 1\). This is nonphysical, as the observed value of $\Omega_b \approx 0.04$. Although $\omega_{eff} = 0$, and the eigen values $(-1, 3/2, 3\beta - 3/2)$ show a saddle point in this case, we can neglect A. 
	
	For \(x = 1 \) represents a complete dominance of the kinetic energy. The critical point is $B=(1,0,0)$, for which $\Omega_b =0$, $\omega_{eff} = (1 - 2 \beta)$ and the eigenvalues $(3(1 - 2 \beta), 2(1 - 3 \beta), 3(1 - \beta) - \sqrt{3/2} \lambda)$. This represents a decelerating phase if $\omega_{eff} > -1/3 \implies \beta < 2/3$. For $1/2 < \beta < 2/3$, the eigenvalues represent a saddle point if $\lambda < \sqrt{6}(1 - \beta)$. For $\beta = 1/2$, the EoS is $\omega_{eff} =0$ and the eigenvalues always provide a saddle point, which clearly represents a matter dominated era. For $1/3 < \beta < 1/2$, we have $ \omega_{eff} \  \epsilon \ (0, 1/3)$ and the eigenvalues provide a saddle point. However, if $\beta< 1/3$ the eigenvalues represent a saddle point if $\lambda > \sqrt{6}(1 - \beta)$. 
	
	For $\beta = 1/2$, (\ref{constrain}) provides $x^2 + \Omega_b =1$. The critical point is $C=(c,0,0)$ where $\Omega_b = 1 - c^2$. The EoS is $\omega_{eff} = 0$ and the eigenvalues are $(-1 , 0, 3/2 - c \lambda \sqrt{3/2})$. The point $C$ converges to the above points $A,B$ for $c\rightarrow 0,1$, respectively, without any new physics.

	\subsubsection{Case 2: \(y = 0\) and \(\Omega_r \neq 0\)}
	
	From (\ref{con-3}), we have the following
	\begin{equation} \left( x^2 + \frac{\Omega_r}{3} - 2 \beta x^2 \right) = 1/3 ,\label{con-1}
	\end{equation}
	which substituting into (\ref{x-3}) provides 
	\[ x \left( -3 + 3 \beta + \frac{3}{2} (1+ 1/3) \right) = 0\]
	\[ \implies x =0, \ \beta = 1/3. \]
	For $x=0$, (\ref{con-1}) provides \( \Omega_r = 1 \implies \Omega_b=0\), we have the radiation dominated critical point $D=(0,0,1)$ with $\omega_{eff} = 1/3$. The eigenvalues $(1,2, 3 \beta -1)$ provides is a saddle point if $\beta < 1/3$ and an unstable node if $\beta \geq 1/3$ 
	
	For $\beta = 1/3$, (\ref{con-1}) reduced to the Friedmann constraint \( x^2 + \Omega_r = 1 \implies \Omega_b = 0 \). The critical point is $E=(c,0, 1 - c^2)$, the eigenvalues are $(0,1, 2 - c \lambda \sqrt{3/2})$ and the EoS is $\omega_{eff} = 1/3$. The point $E$ converges to the above points $B,D$ for $c \rightarrow 1,0$, respectively.   
	
	\subsubsection{Case 3: \(y \neq 0\) and \(\Omega_r = 0\)}
	
	In this case, we solve for the variables $(x,y)$ for a given value of $\Omega_b$. From (\ref{con-2}), we have
	\begin{equation}
		1+ x^2 - y^2  - 2 \beta x^2  = \lambda \sqrt{\frac{2}{3}} x. \label{x4}   
	\end{equation}
	Substituting which into (\ref{x-3}) provides
	\begin{eqnarray}
		- 3 x + 3 \beta x + \sqrt{\frac{3}{2}} \lambda ( x^2 + y^2 )  = 0 \label{x6}.
	\end{eqnarray}
	On the other hand, we have the constraint $x^2 + y^2 + \Omega_b = 1$. For a given $\Omega_b$, the above equations (\ref{x4}, \ref{x6}) produces a scaling solution of the form
	\begin{equation}
		x = \frac{(1 - \Omega_b) \times \lambda}{\sqrt{6} (1 - \beta)} , \ y = \sqrt{1 - \frac{(1 - \Omega_b^2)) \times \lambda^2}{6 (1 - \beta)^2} - \Omega_b} \label{scale-sol}.
	\end{equation}
	The EoS with the above solution reads $\omega_{eff} = -1 + \frac{(1 - \Omega_b)^2 \lambda^2}{3 (1 - \beta)} + \Omega_b$, whereas the eigenvalues are calculated in Appendix - \ref{eigen}. Clearly, for $\Omega_b \rightarrow 0$ and $\lambda \rightarrow \sqrt{6} (1 - \beta)$, the above solution approaches the matter dominated critical point $B$ with the eigenvalues $(2 - 6 \beta,3 - 6 \beta,0)$ and the EoS $\omega_{eff} \approx (1 - 2 \beta)$. On the other hand, for a very small value of $\lambda$, the above solution approaches the point $\approx (0,1,0)$ with eigenvalues $\approx (-4,-3, -3(1 - \beta))$ and EoS $\omega_{eff} \rightarrow -1$. This represents the late time cosmic acceleration (a stable attractor with all `-' eigenvalues) if $\beta < 1$ and the universe enters into the de-sitter phase.  
	
	\subsubsection{Case 4: \(y \neq 0\) and \(\Omega_r \neq 0\)} 
	
	In this case, we have the above three equations (\ref{x-3}, \ref{y-3}, \ref{omega-3}) with the Friedmann constraint (\ref{constrain}), which is unable to provide any consistent solution and thus can be neglected. \\
	
	In the above cases, we found that the points $B(0,0,1),D(1,0,0)$ represent radiation and matter dominated eras, respectively, whereas the scaling solution justifies both the matter dominated and late time de-sitter phase. It can be realized that, for $\beta>1/3$, the radiation dominated phase (for $B(0,0,1)$) becomes an unstable point, but the latter phases can be clearly explained for $1/3 < \beta < 2/3$. In the scaling solution evaluated above (\ref{scale-sol}), one can in principle use the physical value of $\Omega_b \approx 0.04$ instead of its vanishing value, but the above analysis will remain unchanged for some suitable choices of $\lambda$. Below we represent a plot for the cosmological evolution from radiation domination to the late time accelerating phase for the above case.  
	\begin{figure}[H]
		\centering
		\includegraphics[width=0.6\textwidth, height=6cm]{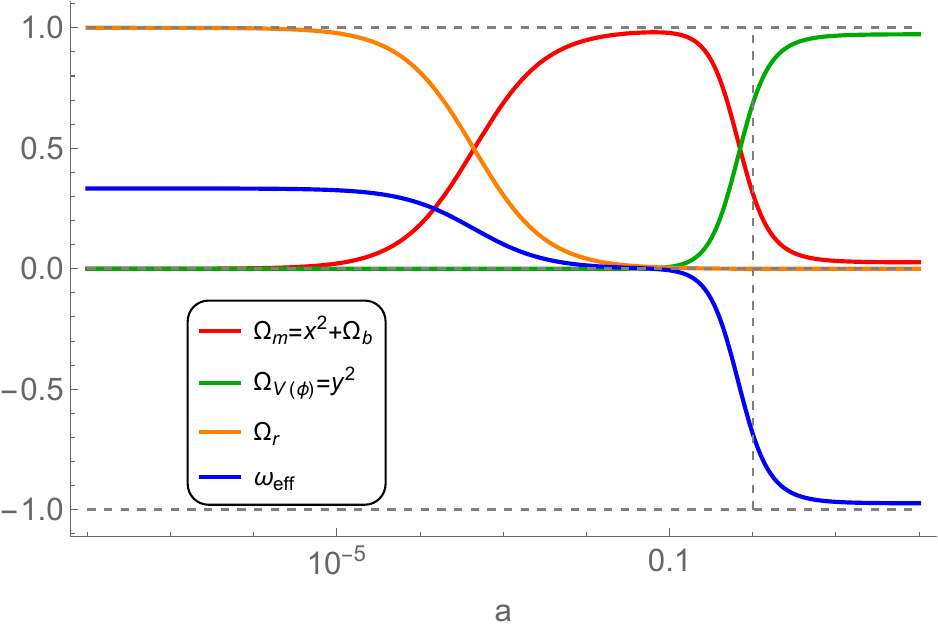}
		\caption{The evolution of cosmological eras have been shown from very past $a(t) = 10^{-8}$ to some future values $a(t)=10^2$. To represent the contribution of matter we considered the total matter density parameter as $\Omega_m = x^2 + \Omega_b$. The late time de - sitter phase is considered due to the domination of scalar field potential term $y^2$. For the above plot, the values of $\beta, \lambda$ are taken to be $1/2$ and $0.2$, respectively. The above plot clearly explains the cosmological eras.}
		\label{Pic-1}
	\end{figure}
	
	\subsection{Phase space analysis with \(\lambda = 0\)}
	
	This is the case for $V(\phi)$ a constant, so that $\lambda =0$. The analysis of the critical points is as follows: 
	
	\subsubsection{Case 1: \(y = 0\) and \(\Omega_r = 0\):}
	
	The critical point $(0,0,0)$ has the same eigenvalues as previously, which is nonphysical. However, the eigenvalues for the critical point $(1,0,0)$ are now $(3(1 - 2 \beta), 2(1 - 3 \beta), 3(1 - \beta) )$  with the EoS $\omega_{eff} = (1 - 2 \beta)$. Clearly, the point represents matter dominated phase if $\beta = 1/2$ and a decelerating phase (a saddle point) if $1/3< \beta < 2/3$, for which the EoS ranges between $(1/3 , -1/3)$. The analysis of the critical point $(c,0,0)$ with $\Omega_b = 1 - c^2$ (for $\beta = 1/2$) is the same as before, but with $\lambda =0$. 
	
	\subsubsection{Case 2: \(y = 0\) and \(\Omega_r \neq 0\):}
	
	Once again, the analysis is the same as before for the critical point $(0,0,1)$. On the other hand, the eigenvalues for the critical point $(c,0, 1 - c^2)$ are now $(0,1,2)$, which is clearly unstable for $\beta = 1/3$ and thus can be neglected.  
	
	\subsubsection{Case 3: \(y \neq 0\) and \(\Omega_r = 0\)}
	
	In this case, a critical point is $(0,1,0)$, for which the eigenvalues are $(-4, -3, -3(1 - \beta))$ and the effective equation of state is $\omega_{eff} = -1$. This point clearly represents a late-time accelerated phase and is a stable attractor if $\beta < 1$. Another critical point arises for $\beta = 1$, which yields the condition $x^2 + y^2 = 1$. This leads to a scaling solution $x = c$, $y = \sqrt{1 - c^2}$, corresponding to the critical point $(c, \sqrt{1 - c^2}, 0)$. The eigenvalues for this point are $(-4, -3, 0)$, with the equation of state $\omega_{eff} = -1$. One can see that as $c \rightarrow 1$, this point approaches $(1,0,0)$, and as $c \rightarrow 0$, it approaches $(0,1,0)$. However, the value $\beta = 1 $ lies outside the domain of physical interest. The fourth case $y \neq0, \ \Omega_r \neq 0$ is the same case as previously and thus can be neglected again.
	
	Therefore, we again arrive at the same constraint on $\beta$, namely $1/3 < \beta < 2/3$, to consistently describe the full cosmological evolution, from the radiation era to the late-time accelerated phase. As above, the evolution is shown by the following plot.   
	\begin{figure}[H]
		\centering
		\includegraphics[width=0.6\textwidth, height=6cm]{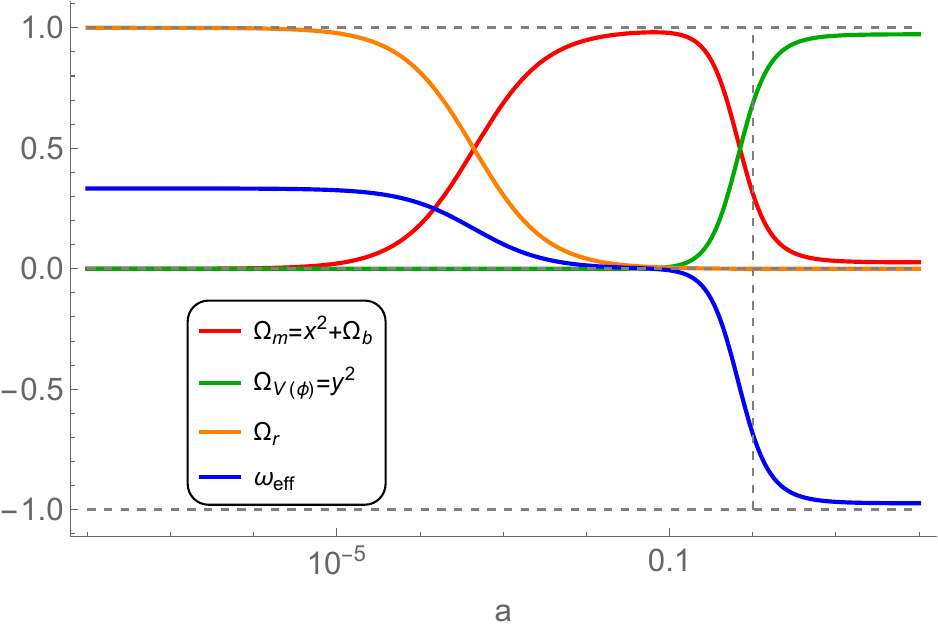}
		\caption{Once again we have considered the values of $a(t)$ from $10^{-8}$ to $10^2$. For the above plot, the values of $\beta$ is taken to be $1/2$.  The above plot clearly explains the cosmological eras similar to the above case.}
		\label{Pic-2}
	\end{figure}

	\section{A comparison with observational data}
	\label{V}
	
	The acceleration equation (\ref{acc-eqn}) can be written in the following form as
	\begin{equation}
		\frac{dH}{dN} = \frac{\dot{H}}{H} = -\frac{3}{2} \left(1 + \omega_{\text{eff}} \right) H \label{H-eq}   
	\end{equation}
	where, $\omega_{eff}$ is given by (\ref{omega-eff}). For a given value of $N$, the above equation (\ref{H-eq}) produces the solution 
	\begin{equation}
		H(N) = H_0 \ e^{-\frac{3}{2} \int_0^N [1 + \omega_{eff}(N')] dN'}, \label{H-sol}
	\end{equation}
	where, $H_0 = H(0)$ is the present value of Hubble parameter. In terms of red-shift $z$ ($z = e^{-N} - 1 $), the above Hubble parameter $H$ provides the luminosity distance 
	\begin{equation}
		d_L(z) = (1 + z) \int_0^z \frac{dz'}{H(z')},    \label{lum} 
	\end{equation}
	applying which one can calculate the distance modulus (assuming \( c = 299792.458\,\mathrm{km/s} \) ) as
	\begin{equation}
		\mu(z) = 5 \log_{10}(d_L(z)) + 25. \label{mu-z}    
	\end{equation}
	However, for practical purpose people often uses the dimensionless variable \( E(z) = \frac{H(z)}{H_0} \) to compare theory with observations. With this, the luminosity distance becomes
	\begin{equation}
		d_L(z) = \frac{(1+z)}{H_0} \int_0^z \frac{dz'}{E(z')}.    \label{lum-1} 
	\end{equation}
	We will use (\ref{mu-z}, \ref{lum-1}) written with the dimensionless Hubble parameter $E(z)$ to compare our model with observations. Especially, to study the evolution of Hubble parameter we uses the data provided in \cite{farooq}, whereas for the comparison of distance modulus we uses the recent Pantheon Supernova Data \cite{PantheonPlus2022}. 
	
	\begin{figure}[ht]
		\centering
		\subfigure[For $\lambda=0$.]{
			\includegraphics[width=0.45\textwidth]{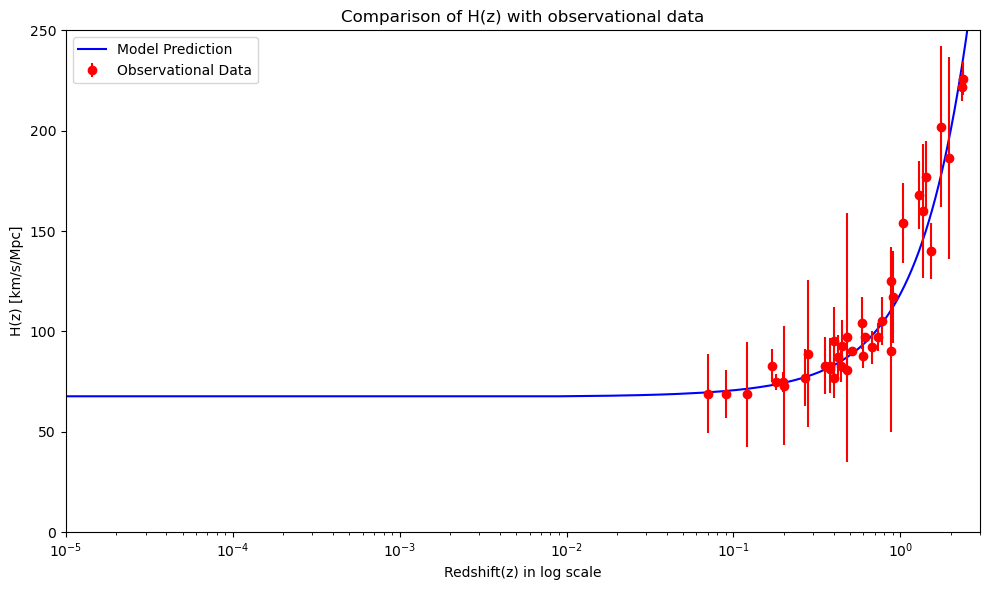}
			\label{fig:sub1}
		}
		\subfigure[For $\lambda=0.2$.]{
			\includegraphics[width=0.45\textwidth]{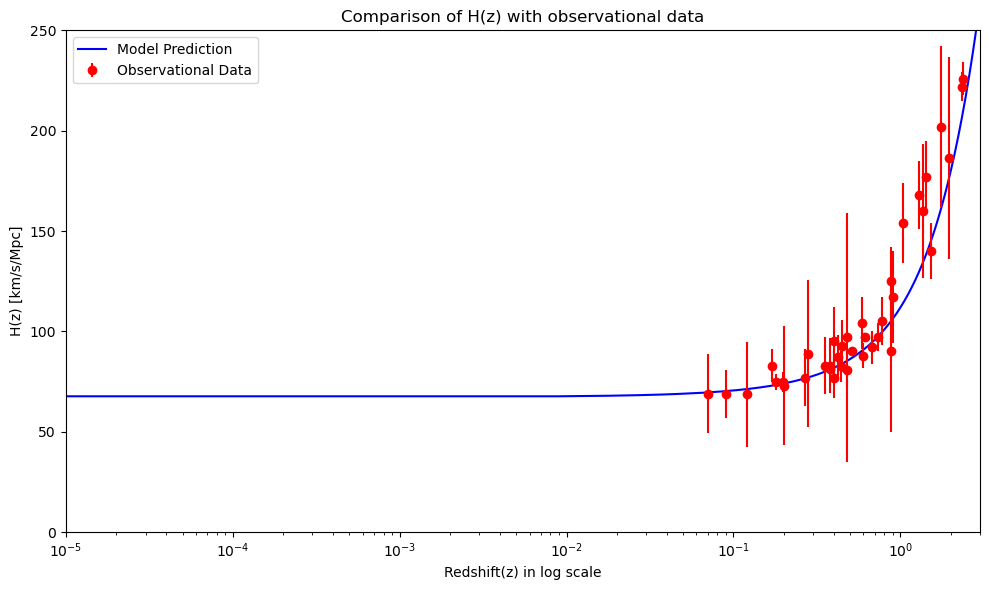}
			\label{fig:sub2}
		}
		\caption{Comparison of the model-predicted Hubble parameter $H(z)$ with observational data  given in \cite{farooq}. The solid blue line represents the evolution of $H(z)$ as obtained from solving (\ref{H-sol}) for the particle creation rate $\Gamma = 3 \beta H$ with $\beta = 0.5$, whereas $\lambda=0$ and $\lambda = 0.2$ correspond to the constant and exponential potentials, respectively. The red points with error bars correspond to the observational estimates of the Hubble parameter. The above plot shows that the values of Hubble parameter predicted from our model is closely connected to its observed values.}
		\label{fig:Hz_comparison}
	\end{figure}
	\begin{figure}[ht]
		\centering
		\subfigure[For $\lambda=0$.]{
			\includegraphics[width=0.45\textwidth]{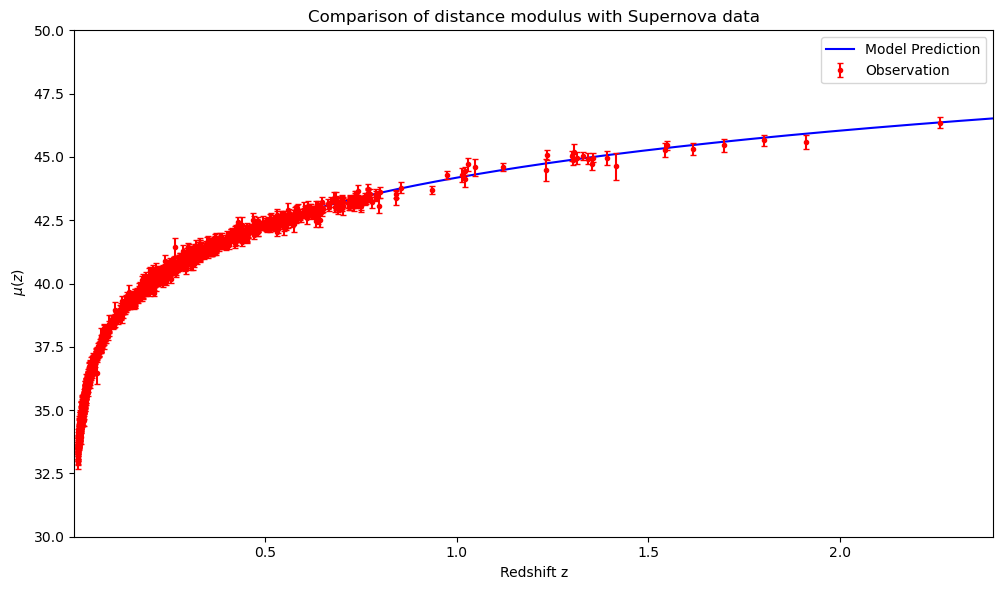}
			\label{fig:sub1}
		}
		\subfigure[For $\lambda=0.2$.]{
			\includegraphics[width=0.45\textwidth]{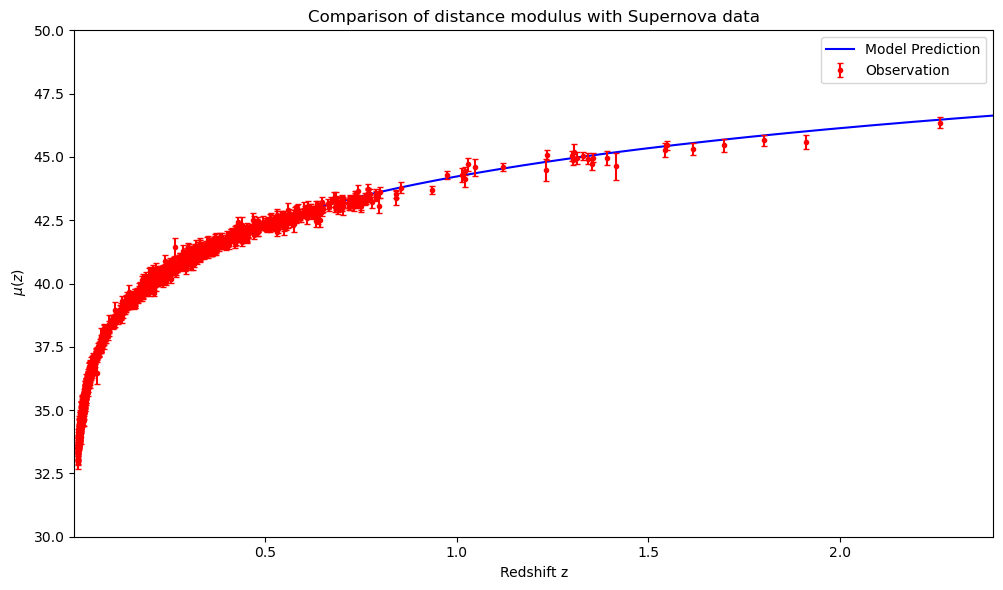}
			\label{fig:sub2}
		}
		\caption{ Comparison of distance modulus $\mu(z)$ as a function of redshift between the particle creation rate $\Gamma = 3 \beta H$ with $\beta = 0.5$ (blue curve) and binned observational data from the Pantheon supernova sample \cite{PantheonPlus2022}(red points with error bars). As previously, $\lambda=0, \ 0.2$ corresponds to constant and exponential potentials. The plot above shows a very good agreement of the observed distance modulus with the theory.  }
		\label{fig:mu_vs_z}
	\end{figure}
	
	From Fig.~\ref{fig:Hz_comparison}, one can realize that our model predicts a late-time expansion history that aligns well with observed data over the redshift range $0 < z < 2$, with some small deviations remaining with observational uncertainties. This result shows that the action-derived particle creation model with $\Gamma = 3 \beta H$ can efficiently reproduce the expansion history of the universe from the radiation dominated era to the late time cosmic accelerating phase. On the other hand, the distance modulus is compared in Fig.~\ref{fig:mu_vs_z}. Comparison with the binned Pantheon supernova dataset shows that the particle creation model closely follows the observed trend within the redshift range $0 < z < 3$. This agreement again suggests that the particle creation model with the chosen parameters can successfully reproduce late-time cosmic acceleration, consistent with supernova data.

	\section{Summary and Conclusions}
	\label{discuss}
	
	The main purpose of the present work is to investigate whether the theory of cosmological adiabatic particle production, an alternative to explaining the dynamical phases of the universe, follows from a valid Lagrangian description. Starting from a general form of action with an arbitrary coupling between metric and scalar field sectors, we have computed the general form of the field equations. We have restricted the modified term of the Lagrangian by comparing the field equations with the known Friedmann, acceleration, and continuity equations that people use to study in the case of particle creation formulations. We have found that the form of the particle creation model that can be achieved from the derived Lagrangian is $\Gamma = \dot a \times h(a)$. More specifically, if the particle creation pressure is considered as a coupling between the metric and the scalar field (where the kinetic term represents domination of matter) of the following form $p_c = \Gamma/ (3H) \times \dot \phi^2$, then only the model (\ref{model}) has an action description. Thereafter, we have approached to study the phase-space stability analysis with our model. We have formed a complete dynamical system including evolution of our model $\Gamma = \dot a \ h(a)$ and $\Lambda = V V_{\phi\phi}/ V_\phi^2$ (and their higher order evolutions) as additional variables. The intension were to study whether consideration of higher order evolutions of those variables can produce any critical points. We have found that the existence of critical points further restricts the particle creation model to $\Gamma = \dot a \times c_1 / a = 3 \beta H$, whereas the scalar field potential is restricted to $V(\phi) = c_1 , \ c_1 e^{c_2 \phi}$. For both potentials, we have shown that our model is capable of explaining the cosmic evolution from radiation dominated era to the late-time cosmic acceleration if $1/3 < \beta < 2/3$ and for some suitable choice of $\lambda$. The evolution of those eras are presented in Figures -\ref{Pic-1} \& \ref{Pic-2} for the cases $\lambda \neq 0, \ 0$ respectively. To test the observational relevance of our particle creation model, we numerically calculated evolutions of Hubble parameter and distance modulus starting from radiation to present-day cosmological era. The resulting Hubble function $H(z)$ was compared with observational data proposed in \cite{farooq}, whereas for the distance modulus we used Pantheon Type Ia supernova data \cite{PantheonPlus2022}. Our analysis shows that the model, with $\Gamma = 3 \beta H $, is capable of reproducing the observed expansion history of the universe \( 0< z \lesssim 3 \) as an alternative to $\Lambda$CDM model. Fine-tuning of $\beta$ and $\lambda$ may further improve the fit or allow tighter constraints on model. However, the above result supports the viability of an action-based formulation of gravitationally induced particle production in explaining late-time cosmic acceleration.
	
	In the stability analysis section, we came to the forms of $h(a)$ and $ \lambda$ studying their compatibility with higher order evolutions of those variables. This differs from the conventional approach of considering $\Lambda = \Lambda(\lambda)$ in (\ref{lambda}), allowing the investigation of alternative potentials. This perspective might be applied to our model $h(a)$ and could potentially offer valuable insights.   
	
	{\appendix
		
		\section{Eigen values}\label{eigen}
		
		The eigenvalues are 
		\begin{eqnarray}
			&& \lambda_1 = -4 + \frac{\lambda^2}{(1 - \beta)} + 3 \Omega_b - \frac{2\lambda^2 \Omega_b}{1- \beta} + \frac{\lambda^2 \Omega_b^2}{1- \beta}, \nonumber \\
			&& \lambda_{2,3} = -3 + \frac{3 \beta}{2} + \frac{3 \lambda^2}{4 ( 1 - \beta )} + ( 3 - \frac{7\lambda^2}{4(1 - \beta)} ) \Omega_b + \frac{\lambda^2 \Omega_b^2}{1 - \beta} \pm \frac{1}{4} \sqrt{a_0 + a_1 \Omega_b + a_2 \omega_b^2 + a_3 \Omega_b^3 + a_4 \omega_b^4} \nonumber.
		\end{eqnarray}
		where, $a_0 = \frac{(\lambda^2 - 6 \beta ( 1 - \beta))^2}{(1 - \beta)^2 }, a_1 = - ( 72 \beta + \frac{12 (3 - 7 \beta ) \lambda^2}{1 - \beta} - \frac{2 \lambda^4}{(1 - \beta)^2}), a_2 = (36 + \frac{12 (1 - 6 \beta ) \lambda^2}{1 - \beta} + \frac{3 \lambda^4}{(1 - \beta)^2}), a_3 = \frac{4 \lambda^2 (\lambda^2 -6 + 6 \beta)}{(1 - \beta)^2}, a_4 = \frac{4 \lambda^4}{(1 - \beta)^2}$.    
	}

	\section{Acknowledgments}
	The Fellowship of S. Pramanik is sponsored by Ariel University. I would like to thank Prof. M. Lewkowicz and Prof. M. Schiffer for helpful discussions.
	

\end{document}